\documentclass{article}
\usepackage{spconf,amsmath,graphicx,hyperref}
\usepackage{amssymb}

\usepackage{booktabs}
\usepackage{tabularx}
\usepackage{threeparttable}
\usepackage{siunitx}
\usepackage{array}
\usepackage{ragged2e}

\providecommand{\checkmark}{\ensuremath{\surd}}

\usepackage{tikz}
\usetikzlibrary{arrows.meta,positioning,fit,backgrounds}

\newcolumntype{L}[1]{>{\RaggedRight\arraybackslash}p{#1}}

\title{DJustify: Reason-and-Verify DJ Planning for Pop Transitions and Story-Driven Sets}

\name{\begin{tabular}{@{}c@{}}Wei-Han Hsu$^{1,2\ast}$, Li-Jie Lin$^{3\ast}$, Po-Hsuan Lai$^{3}$, Po-Hsiang Huang$^{4}$,\\ Jeng-Yue Liu$^{5}$, Li Su$^{2}$, Yi-Hsuan Yang$^{3,4,6}$\end{tabular}\thanks{$^{\ast}$Equal contribution.}}
\address{$^{1}$Data Science Degree Program, National Taiwan University and Academia Sinica, Taipei, Taiwan\\$^{2}$Institute of Information Science, Academia Sinica, Taipei, Taiwan\\$^{3}$National Taiwan University, Taipei \quad $^{4}$Taiwan AI Labs, Taipei \quad $^{5}$Carnegie Mellon University, USA\\$^{6}$AI Center of Research Excellence (AI-CoRE), National Taiwan University, Taipei, Taiwan}

\begin{document}
\ninept
\maketitle
\providecommand{\checkmark}{\ensuremath{\surd}}
\begin{abstract}
Automatic DJ systems mostly target electronic dance music, whose regular structure makes rule-based mixing relatively easier. Vocal-rich popular music is harder: which track to play next, where to cue it, and which transition to use are tied together, so fixed rules handle them poorly. We present DJustify, a pop AI-DJ that puts an LLM in a reason--propose--verify--repair loop and lets it choose the next track, cue points, and mixing technique together. The LLM samples several transition plans; an external acoustic checker then scores the rendered audio and triggers repair when a check fails. Sampling keeps several candidate transitions per track pair instead of one fixed mix. In blind listening tests our transitions beat Mosaikbox, a reimplemented crossfade, and a cue baseline that reuses the crossfade's cues; that last comparison shows most of the margin comes from cue placement. The same planner also picks and orders a set along an emotion arc: this improves story coherence over a no-story ablation, and within a set story and transition ratings show no measurable gap from a human DJ. The checker mainly adds reliability, not higher average quality: it repairs rare but audible errors such as cut vocals, while the planner stays non-deterministic.
\end{abstract}

\begin{keywords}
Automatic DJ mixing, large language models, rendered-audio verification, output diversity, story-driven DJ sets
\end{keywords}

\section{Introduction}
\label{sec:intro}


A DJ keeps the music playing without a break, working two decks so that each song hands over smoothly to the next; done well, the listener barely notices the join. Automatic DJ systems take on the same choices: which track to play next, where each track enters and leaves the mix, and how the two are blended, so that a set plays as one continuous stream~\cite{bittner2017,vande2018raw}. Most such systems target electronic dance music (EDM), where regular phrase structure, instrumental intros and outros, and a steady beat grid make transitions easy for fixed rules. Vocal-rich pop is harder. Pop songs vary in functional structure---the arrangement of verses, choruses, and bridges~\cite{kim2023allin1}---carry dense, prominent vocals, and span many styles, so transition quality depends not only on how two tracks are mixed but also on which track is chosen, where each is cued, and which technique is used. These decisions are coupled: a technique that works for one pair of cue points fails for another, and changing the next track changes which cues and techniques are feasible at all.


Rule-based systems handle these decisions in different ways. Bittner et al.~\cite{bittner2017} optimize playlist order and crossfade points; Vande Veire and De Bie~\cite{vande2018raw} pick structural cue points and transition types, with vocal-clash detection, for drum and bass; Mosaikbox~\cite{sowula2024mosaikbox} separates stems to suppress clashes during the blend. Each relies on hand-set rules, and none revises its choice after hearing the rendered result. A rule that optimizes these decisions independently and deterministically cannot capture their coupling, and it returns a single answer, excluding other valid choices. What is needed is context-dependent reasoning: reading both tracks' structure, vocal activity, and candidate cues together and proposing the next track, its cue points, and the technique as one plan. A large language model (LLM) can do this, and sampling it gives several plans for the same pair rather than a single answer that was already fixed in advance.


A text-only LLM cannot check its own work, however: it cannot hear the mix it plans, and without external feedback its self-checks are unreliable~\cite{huang2024,valmeekam2023,stechly2025}. We therefore close a reason--propose--verify--repair loop with rendered audio. Given the outgoing track and a candidate pool, the LLM proposes an incoming track, a cue-out on the outgoing and a cue-in on the incoming track, an overlap length, and a transition technique; the plan is rendered, an external checker of deterministic acoustic gates reports which gate failed and what to change, and the planner repairs the plan. Rules act as a \emph{verifier} that bounds the set of valid transitions, not a \emph{decider} that fixes the output, and the same loop scales from one transition to a story-driven set.\footnote{Open-source code and audio demos: \url{https://github.com/ddman1101/DJustify} and \url{https://jamie930625.github.io/djustify/}.}


Our contributions are threefold: (i)~a non-deterministic LLM DJ planner that samples multiple distinct transitions per track pair, each checked and repaired on the rendered audio (\emph{constrained diversity}); (ii)~a transition system for vocal-rich pop---a setting prior Auto-DJ work largely skips---that significantly outperforms the closest prior automatic systems, including a Spotify-style crossfade, with a controlled cue baseline isolating the importance of cue-point selection; and (iii)~to our knowledge the first DJ \emph{mixing} system whose rendered transitions follow an emotional arc, with story and transition ratings showing no detectable gap from a human DJ in a blind listening test.

\begin{figure*}[t!]
\centering
\includegraphics[width=0.98\textwidth]{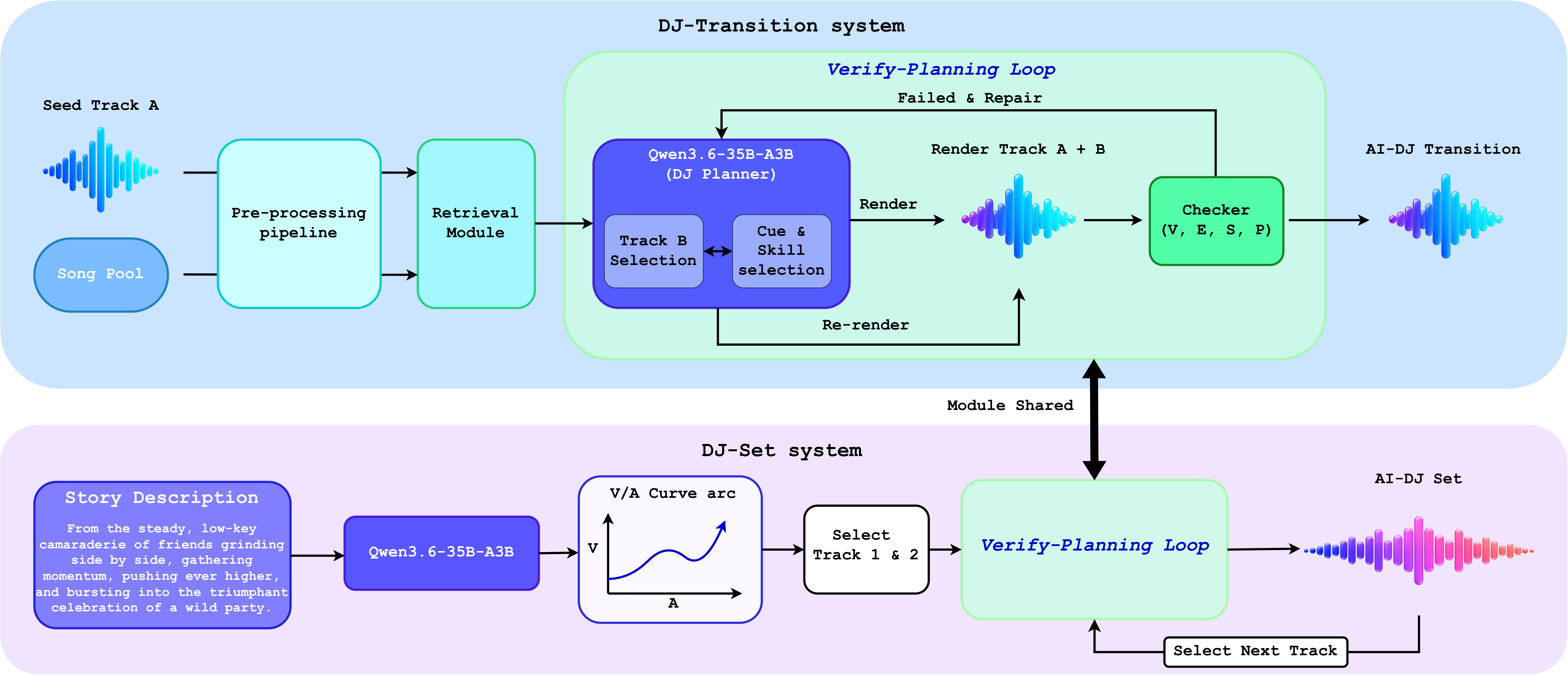}
\caption{
AI-DJ overview. \textbf{Top:} a planner jointly selects the next track, cue points, and technique; the rendered transition is scored by acoustic checks (vocal/energy/structure/pairing) and failures drive repair. \textbf{Bottom:} a story prompt guides track selection and ordering into a valence--arousal arc, invoking the same loop to build a set.
}
\label{fig:overview}
\end{figure*}

\section{Related Work}
\label{sec:related_work}

Related work falls into two lines: automatic DJ mixing on one side, and LLM reasoning, self-correction, and music agents on the other.

Automatic DJ systems assemble a continuous mix by ordering tracks and synthesizing a transition between successive pairs. Early systems sequence by tempo and energy~\cite{cliff2000hang} or pick mix points by beat and latent-topic similarity~\cite{hirai2016}; more complete pipelines cast cue selection and beat-matched blending as constrained optimization~\cite{bittner2017,vande2018raw}, and Mosaikbox adds harmonic compatibility and vocal separation~\cite{sowula2024mosaikbox}. Learning-based methods instead generate the transition directly: DJtransGAN learns cross-fader and EQ curves adversarially~\cite{paul2022}, and DJ-AI sequences a playlist and synthesizes transitions with a generative model~\cite{kinay2025dj}. Two limitations recur. First, they are built and evaluated almost only on EDM; vocals are handled by separating stems~\cite{sowula2024mosaikbox} or by detecting overlapping vocals~\cite{vande2018raw}, not by cue timing that keeps a sung phrase intact, and quality is judged by beat and harmony rather than by whether a sung line survives. Second, selection, cueing, and synthesis are optimized \emph{in isolation}, pair by pair, so the play order has no planned emotional shape. We target pop, join selection, cueing, and synthesis in one reason--verify loop, and plan the play order around an emotional arc, which prior Auto-DJ mixing systems do not do. We do not compare against Vande Veire's full pipeline, which is built for drum and bass.

On the LLM side, chain-of-thought prompting improves reasoning~\cite{wei2022}, though its gains concentrate on symbolic and mathematical tasks~\cite{sprague2025}. A parallel line lets models revise their own output~\cite{madaan2023}, but without external feedback such self-generated critique often fails to improve or even degrades the answer~\cite{huang2024,valmeekam2023}. LLMs have also driven music agents and composition~\cite{yu2023musicagent,zhang2023}; closest to us, Chang et al.~\cite{chang2025cot} use chain-of-thought to fuse MIR tool outputs for chord recognition, but over symbolic labels rather than rendered audio.

\section{Method}
\label{sec:method}

Figure~\ref{fig:overview} gives an overview. The system operates at two levels that share one verified-planning loop: a single transition, and a full story-driven set. Both levels are driven by a single instruction-tuned LLM (Qwen3.6-35B-A3B~\cite{team2026qwen3}): the same model plans a transition, selects the next track, and orders a set, differing only in the task prompt. Everything runs locally on one 48\,GB NVIDIA RTX 6000 Ada GPU with a 4-bit-quantized LLM; no closed-source APIs are called.

\subsection{Song Preprocessing}
\label{ssec:song_pre}
Off-the-shelf estimators give each track a beat/downbeat grid, a structural segmentation, and chords~\cite{kim2023allin1,crema}. Two features drive cue placement and the checker: per-frame singing-voice activity from a CRNN we train for singing-voice detection~\cite{svd_schluter,svd_leglaive} on the Jamendo corpus~\cite{jamendo_dataset}; and a valence/arousal label from a regression baseline over MuQ-MuLan embeddings~\cite{muq}, refined in context by an audio-language model reading the audio and its transcribed lyrics~\cite{qwen3omni,whisper}. Cue \emph{anchors}---phrase ends, section boundaries, harmonic restarts, and low-frequency onsets---restrict cues to musically strong, on-grid points and limit where the planner can cue.

\subsection{Verified DJ Transition Planning}
\label{ssec:trans}
Given the outgoing track $A$, a retrieval module supplies a candidate pool, and the planner jointly selects the incoming track $B$ together with a cue-out on $A$ and a cue-in on $B$---each chosen from a precomputed shortlist of candidate cues at the anchors of \S\ref{ssec:song_pre} (a \emph{cue card})---an overlap length, and a mixing technique. Techniques range over a continuum of overlap~\cite{morse2018}, from \emph{blend} moves (long overlap, e.g. a chorus riding over an outgoing outro) to \emph{cut} moves (abrupt handover with an effect: echo, brake, backspin, filter). Reasoning corresponds to the model's explicit thinking mode: with it disabled the LLM emits the structured plan directly, whereas enabling it lets the model deliberate over structure, vocal activity, cue anchors, and technique before committing. Because the symbolic anchors are only resolved to on-grid seconds at render time, a plan valid on paper can still fail once rendered, which is why verification runs on the rendered audio. We use deterministic acoustic gates rather than an audio-language-model critic, which can rate audio without reliably perceiving it~\cite{zang2025listening}.

The checker is a set of deterministic acoustic gates over the \emph{rendered} signal, in four families (Table~\ref{tab:checker}). A failing gate returns an \emph{actionable} repair---re-cue one phrase later, shorten the overlap, switch technique---not just a rejection, which the planner acts on for up to four rounds; a plan still failing after that is returned flagged, not resampled. Toggling the two mechanisms gives three ablation variants: \textbf{Ours} (reasoning $+$ checker-driven repair), \textbf{Ours$-$C} (reasoning only), and \textbf{Ours$-$C$-$R} (neither).

\begin{table}[t]
\centering
\setlength{\belowcaptionskip}{5pt}
\caption{The checker: rendered-audio gates and corrective actions.}
\vspace{5pt}
\label{tab:checker}
\fontsize{9}{10.5}\selectfont
\setlength{\tabcolsep}{4.5pt}
\renewcommand{\arraystretch}{1.18}
\begin{tabular}{@{}l p{6.9cm}@{}}
\toprule
\textbf{Gate} & \textbf{Detected failure $\rightarrow$ corrective action} \\
\midrule
\textbf{Vocal} & Cut mid-phrase $\rightarrow$ shift the cue to the phrase end \\
\textbf{Vocal} & Vocals collide in the overlap $\rightarrow$ shorten it or cut \\
\textbf{Energy} & Silent gap or energy hole $\rightarrow$ adjust length or technique \\
\textbf{Structure} & Early cut in $A$ or late entry in $B$ $\rightarrow$ move the cue \\
\textbf{Structure} & Entry stays in $B$'s intro $\rightarrow$ re-cue to the verse or chorus \\
\textbf{Pairing} & Valence--arousal mismatch $\rightarrow$ re-select track $B$ \\
\bottomrule
\end{tabular}
\end{table}

\subsection{Story-Driven DJ Set}
\label{ssec:set}
To build a set, the same LLM selects and orders tracks into an emotional arc. Starting from a given opening track, it draws from the shared pool the songs whose lyrics, valence/arousal, tempo, and key best realize a rising-and-resolving narrative (opening, build, close), one track per act of the story, under tempo and key compatibility gates. Each resulting junction is then rendered and verified by the transition loop of \S\ref{ssec:trans} with the pairing fixed, and lyric-hook landing points additionally inform cue placement within each transition. We report \textbf{Story} and \textbf{No-story} (an ablation with the same gates and checker but no story, picking each next track by acoustic compatibility alone), which isolates the effect of story planning.

\section{Experiments and Results}
\label{sec:exp}

\subsection{Setup}
\label{ssec:setup}

We run two blind listening tests and an objective ablation. In the \emph{transition test} (Test1), twelve seed tracks are drawn from real DJ sets; for each, every system is given the same fixed track pair, so the comparison isolates cue placement and transition synthesis (next-track selection is exercised in the DJ-set test, Test2, below; with the pair fixed, the pairing gate that would re-select the incoming track is inactive here). Each participant rates, on a 1--10 scale, transitions from our variants (Ours, Ours$-$C, Ours$-$C$-$R), a human DJ, and three baselines: Mosaikbox~\cite{sowula2024mosaikbox} (released implementation), a Spotify-style crossfade we implement following Bittner et al.~\cite{bittner2017} (denoted Bittner'17/Spotify), and a cue baseline, \textbf{Ours (Bittner cues)}, that runs our full pipeline---including the repair loop---but \emph{seeds} its cues from Bittner's detector~\cite{bittner2017} instead of our vocal-aware anchor selection. In both tests each of our stimuli is the planner's output, generated once, with no re-sampling or selection. In the \emph{DJ-set test} (Test2), each participant is given one of three story prompts and compares three-minute sets, rating \emph{story} coherence and \emph{transition} quality separately; conditions are Story, No-story, a human DJ, and the two automatic systems. Story, the crossfade, and Mosaikbox start each set from the same prescribed opening track, while No-story draws its opener at random from tracks with a valid exit anchor, so the Story--No-story contrast also includes the opener. The human condition was produced by a professional DJ: in Test1 they mix the same fixed pair given to every system, while in Test2 they build each set freely, choosing the tracks, cue points, and techniques. All track selection---by our planner, the automatic baselines, and the human DJ---draws from a shared, self-collected pool of 400 full-length tracks, mostly English-language and Mandarin with some Korean and Japanese pop songs, spanning roughly 75--140 BPM and analyzed as in \S\ref{ssec:song_pre}. All stimuli are loudness-normalized and beat-matched where applicable; each participant rates a randomly assigned subset of the trials. We collected $N{=}103$ respondents for Test1 and $N{=}99$ for Test2; 28 of the Test1 respondents report hands-on DJ experience, and we report their ratings separately where relevant. We report mean opinion scores (MOS) with 95\% bootstrap confidence intervals, and test differences with permutation tests: paired for Test1 (each comparison restricted to listeners who rated both conditions), and stratified by story for Test2, where each participant hears one story and rates a few of its conditions.

For the objective ablation, we measure \emph{renderability}---whether a sampled plan can be executed without violating hard timing or effect-window constraints---and output diversity, sampling the planner ten times per fixed pair and counting distinct transitions. Two transitions count as distinct if they differ in technique, cut (cue-out) point, or entry (cue-in) point. Unless noted, the planner samples the transition at temperature $0.6$ and track selection at $1.0$: higher for varied track choices, lower to keep mixes reliable.

\begin{figure}[t]
\centering
\includegraphics[width=0.95\linewidth]{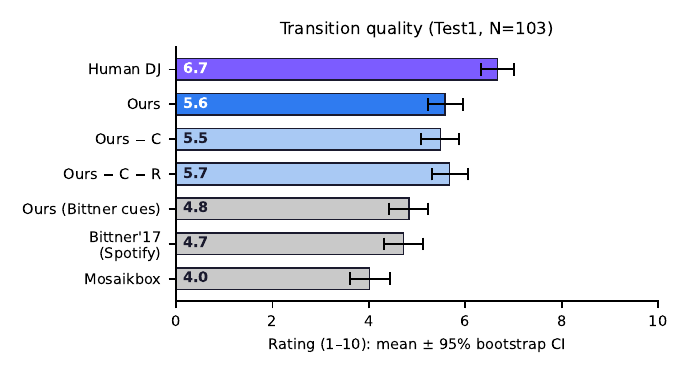}
\caption{Transition quality in Test1 ($N{=}103$). Bars show mean ratings; error bars are 95\% bootstrap confidence intervals. Ours significantly outperforms Mosaikbox, Bittner'17/Spotify, and Ours (Bittner cues); replacing our selected cues with Bittner'17 cues reduces the mean score from 5.6 to 4.8.}
\label{fig:test1}
\end{figure}

\subsection{Transition Quality and Cue Placement}
\label{ssec:transition_results}

Figure~\ref{fig:test1} reports Test1. Mosaikbox is an EDM mixer run here on pop, and Bittner'17/Spotify is our reimplementation of a beat-matched crossfade, not Bittner's full playlist optimizer. Ours is preferred over Mosaikbox (5.6 vs.\ 4.0; paired $\Delta=+1.83$, $p<10^{-4}$), the crossfade (5.6 vs.\ 4.7; $\Delta=+0.89$, $p=.001$), and Ours (Bittner cues) (5.6 vs.\ 4.8; $\Delta=+0.75$, $p=.013$), and stays below the human DJ (5.6 vs.\ 6.7; $\Delta=-1.02$, $p=.0003$); all four comparisons remain significant under Holm correction. Among the 28 DJ-experienced listeners the same pattern holds, and they place the human DJ further ahead ($\Delta{=}{-}1.9$) than the other listeners do.

The comparison with Ours (Bittner cues) isolates the value of our cue selection. It runs the same pipeline and repair loop but seeds its cues from Bittner's detector rather than our vocal-aware anchors; it drops to 4.8, below all three variants (5.7, 5.5, and 5.6 for Ours$-$C$-$R, Ours$-$C, and Ours; $p=.013$ against Ours). Even with repair, seeding from an external detector lands on different, lower-rated cues (matching ours on 1 of 12 pairs) and does not reach our level. For vocal-rich pop, then, transition quality depends on the cue candidates the planner starts from; repair did not close that gap.

\begin{figure}[t]
\centering
\includegraphics[width=0.92\linewidth]{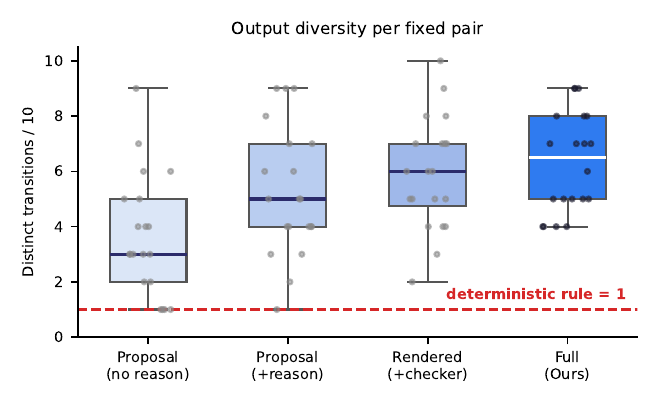}
\caption{
Output diversity over 20 fixed track pairs: for each pair we sample the planner 10 times ($T{=}0.6$) and count \emph{distinct} transitions (differing in technique, cut, or entry). The four settings are proposals without/with reasoning and rendered outputs with the checker (reasoning off) and the full system (Ours). A deterministic rule gives one (dashed line); reasoning raises diversity and repair adds a little.
}
\label{fig:diversity}
\end{figure}

\begin{table}[t]
\centering
\setlength{\belowcaptionskip}{5pt}
\caption{Renderability of planner variants, 100 sampled seeds.}
\vspace{5pt}
\label{tab:renderability}
\fontsize{9}{10.5}\selectfont
\setlength{\tabcolsep}{8pt}
\renewcommand{\arraystretch}{1.15}
\begin{tabular}{@{}lccc@{}}
\toprule
\textbf{Variant} & \textbf{Reasoning} & \textbf{Checker} & \textbf{Renderability} \\
\midrule
Ours$-$C$-$R & -- & -- & 91\% \\
Ours$-$C & $\checkmark$ & -- & 98\% \\
Ours & $\checkmark$ & $\checkmark$ & \textbf{99\%} \\
\bottomrule
\end{tabular}
\end{table}

\begin{figure}[t]
\centering
\includegraphics[width=1\linewidth]{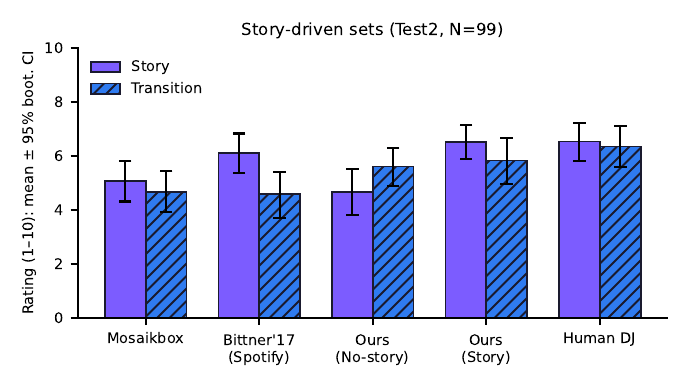}
\caption{
Story-driven set quality (Test2, $N{=}99$), mean story and transition ratings. Story is rated above No-story on \emph{story} ($p{=}.002$) and its transitions beat the crossfade ($p{=}.043$, uncorrected).
}
\label{fig:test2}
\end{figure}

\subsection{Diversity and Reliability under Verification}
\label{ssec:ablation}

Figure~\ref{fig:diversity} compares output diversity across all four combinations of reasoning and checker-driven repair, using ten samples for each of 20 fixed track pairs. The full system returns a median of 6.5 distinct rendered transitions per pair (IQR 5--8), versus one for the deterministic reference. Reasoning increases the median number of distinct proposals from three to five (paired permutation test on the mean difference, $p=.004$). With checker-driven repair, the median count rises from three to six without reasoning ($p=.0003$); with reasoning enabled, it rises from five to 6.5, though not significant ($p=.054$); with reasoning on, repair then adds only a little more.

The no-checker conditions count proposals, whereas the checker conditions count returned rendered outputs. During repair, failed checks prompt changes to cues and techniques rather than rejection. These counts include returned outputs whether or not they pass every gate, showing that multiple alternatives remain after verification.

Table~\ref{tab:renderability} reports execution success over 100 sampled seeds. Renderability rises from 91\% in Ours$-$C$-$R to 98\% in Ours$-$C and 99\% in Ours, with most of the gain coming from reasoning. Most residual failures involve symbolic anchors resolved too close to a track boundary. Renderability measures whether a plan can produce audio, not whether the transition sounds good. The listening test evaluates Ours$-$C$-$R, Ours$-$C, and Ours, adding reasoning and then checker-driven repair; we observe no significant overall difference in their ratings (means 5.7/5.5/5.6; permutation $p>.3$). Checker-driven repair nevertheless corrects specific audible failures. On two seeds, Ours$-$C$-$R cuts an outgoing lead vocal mid-phrase, and Ours$-$C retains these cuts, with DJ-experienced listener ratings of 2.0 and 1.8. Ours removes the cuts, with ratings of 5.0 and 4.8. These are rare rescues, not a measured vocal-error rate; they do not establish an overall perceptual-quality gain. Together, the results support generating multiple alternatives with checking and repair, rather than prescribing one fixed mix, which is the \emph{constrained diversity} of contribution~(i).

\subsection{Story-Driven Sets}
\label{ssec:set_results}

Figure~\ref{fig:test2} tests whether the planner scales from single pairs to complete sets. On transition quality, Story rates above the Bittner'17/Spotify crossfade ($p=.043$, uncorrected) and shows no significant difference from the human DJ ($p=.42$). The set-level parity shows no detectable gap, not that none exists.

The story ratings further isolate set-level planning. Story improves coherence over No-story by roughly 1.9 points ($p=.002$), although the two share the transition engine and are indistinguishable on transition quality ($p=.85$); Story is also indistinguishable from the human DJ on story coherence ($p=.95$). The two draw from the same pool and chose different songs in every set, so the gain comes from selection and ordering together (and the opener policy), not from ordering alone. The comparison with Bittner'17/Spotify does not reach significance for story coherence ($p=.31$).

\section{Conclusion}
\label{sec:conclusion}
We presented a pop AI-DJ built around a reason--propose--verify--repair loop, in which an LLM jointly plans track selection, cue placement, and transition technique, while sampling gives multiple plausible plans for the same context. Blind listening tests show that our transitions remain below a human DJ yet outperform prior automatic systems, while our story-driven sets show no detectable gap from the human DJ in either story or transition ratings. Verification improves reliability rather than average quality---repairing rare, audible vocal-cut failures---while preserving multiple rendered alternatives per track pair, yielding constrained diversity rather than a single deterministic mix. Separating plan generation from rendered-audio verification is thus a practical design for AI-DJ systems. Future work will add audio-language models~\cite{ghosh2026music} as verifiers.





\bibliographystyle{IEEEbib}

\newpage
\vspace{-6pt}
\begin{center}\textbf{ACKNOWLEDGMENTS}\end{center}\vspace{-4pt}
\noindent The work is supported by grants the National Science and Technology Council of Taiwan (NSTC 114-2628-E-002-013-MY3), and the Ministry of Education (MOE) of Taiwan (for Taiwan Centers of Excellence in Artificial Intelligence).

\bibliography{strings,refs}

\end{document}